\documentclass{article}
\usepackage{spconf,amsmath,graphicx,hyperref}
\usepackage{subcaption,multirow}
\usepackage{stfloats}
\usepackage{flafter}
\usepackage{amsmath,amssymb,amsfonts}
\usepackage{booktabs}
\usepackage{makecell}
\usepackage[export]{adjustbox}
\usepackage{enumitem}

\let\oldthebibliography\thebibliography
\renewcommand{\thebibliography}[1]{%
  \oldthebibliography{#1}%
  \small
  \setlength{\itemsep}{-1.0pt}%
}

\title{FAM-Gate: Field-Aware Metadata Attention with Prior-Constrained Vector Gating for Audio Classification}

\name{Xiangyu Jing$^{\dagger}$, Hongqing Liu$^{\dagger, \star}$, Liming Shi$^{\dagger}$}

\address{
$^{\dagger}$ School of Communications and Information Engineering,\\
Chongqing University of Posts and Telecommunications, Chongqing 400065, China\\
$^{\star}$ Corresponding Author: \texttt{hongqingliu@outlook.com}
}

\begin{document}
\maketitle
\ninept

\begin{abstract}
User-generated audio collections frequently contain textual metadata such as titles, tags, and descriptions. Effectively leveraging these complementary fields requires accounting for both field heterogeneity and varying semantic relevance relative to the acoustic content. We propose FAM-Gate, a two-stage framework for metadata-enriched audio classification. In the first stage, field-aware metadata-view attention dynamically aggregates multiple metadata views using audio-guided weighting, explicit field-identity embeddings, a shared identity-initialized projection, and mean-anchored regularized aggregation. In the second stage, a prior-constrained vector gate performs bounded, dimension-wise fusion of audio and metadata representations centered around an acoustic prior. Experiments on BSD10K, BSD35K, and FSD50K show consistent improvements in both hierarchical single-label and multi-label classification. Compared with matched random-view scalar-fusion baselines, FAM-Gate improves the hierarchical F-score by 2.53 and 5.72 percentage points on BSD10K and BSD35K, respectively, while increasing macro mAP and macro F1 by 5.07 and 10.25 percentage points on FSD50K.
\end{abstract}

\begin{keywords}
Audio classification, audio-language models, metadata fusion, field-aware attention, multimodal learning
\end{keywords}

\section{Introduction}

General-purpose sound classification has advanced from hand-crafted features to deep neural representations. Piczak demonstrated that convolutional networks extract discriminative spectro-temporal patterns directly from audio~\cite{piczak2015cnn}, while backbones like PANNs established the transferability of AudioSet pretraining to downstream tasks~\cite{audioset,panns}. Despite these advances, audio-only models often struggle when semantically distinct events share similar acoustic properties or exhibit high intra-class variance.

Text offers a natural source of complementary semantics. SemanticAC aligned audio representations with class-label language embeddings~\cite{semanticac}, while CLAP and LAION-CLAP learned shared audio-text spaces for zero-shot classification~\cite{clap,laionclap}. However, these models typically rely on class-level text (e.g., category names), which lacks recording-specific context.

Sample-level metadata fills this gap. For instance, LRID-Net incorporated titles, album names, and artists for singing-language identification~\cite{lrid}. Moummad and Farrugia leveraged age and sex attributes for respiratory sound pretraining~\cite{moummad2023}, while BTS converted patient and recording metadata into prompts during inference~\cite{bts}. For general audio, Freesound metadata (titles, tags, descriptions) forms the foundation of FSD50K and BSD~\cite{freesound,fsd50k,bsd2024}. Recently, Anastasopoulou et al. concatenated title, tag, and description fields in BSD, fusing the text embedding with CLAP audio features via a scalar gate~\cite{bsd2025}. Generic cross-modal methods like MMTM~\cite{mmtm} and FiLM~\cite{film} also perform feature recalibration, but they neither account for metadata field noise nor constrain fusion around the acoustic prior.

Overall, two key limitations persist in existing approaches:
(1) Titles, tags, and descriptions vary in structure, semantic density, and noise level; simple concatenation or random view selection ignores recording-dependent field relevance.
(2) Scalar gating applies a uniform fusion weight across all feature dimensions, failing to capture channel-wise alignment or conflict between acoustic and semantic evidence.

To resolve these issues, we present Field-Aware Metadata Attention with Prior-Constrained Vector Gating (\textbf{FAM-Gate}). Field-Aware Metadata View Attention (MVA) preserves the pretrained CLAP alignment while dynamically weighting metadata fields using audio context, explicit field identities, and a mean anchor. Subsequently, a Prior-Constrained Vector Gate (PCVG) performs bounded, dimension-wise audio-metadata fusion centered around the acoustic prior. Our main contributions are:
\begin{itemize}[leftmargin=*,noitemsep,topsep=1pt]
    \item An alignment-preserving, field-aware, and mean-anchored mechanism for heterogeneous metadata aggregation.
    \item A prior-constrained vector gate that models dimension-wise cross-modal complementarity and conflict.
    \item A matched five-seed evaluation on BSD10K, BSD35K, and FSD50K, demonstrating consistent gains over matched baselines (+5.72 hF on BSD35K, +10.25 macro-F1 on FSD50K) with only a 0.19\% increase in total parameters.
\end{itemize}

\begin{figure*}[t]
    \centering
    \includegraphics[width=\textwidth]{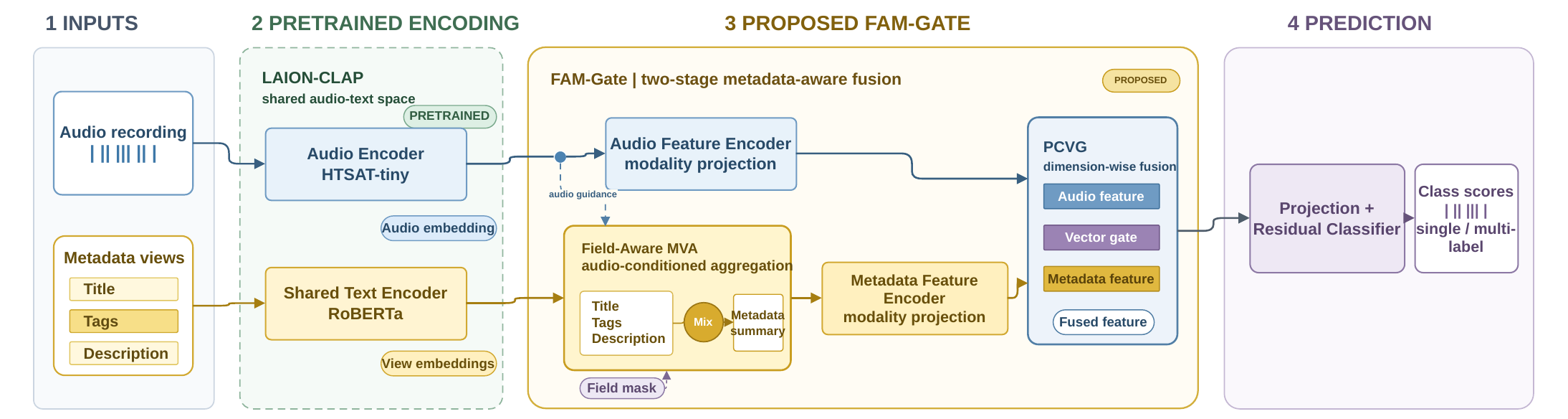}
    \caption{Overall architecture of FAM-Gate. LAION-CLAP embeds the audio and heterogeneous metadata views in a shared space. The audio embedding guides MVA in metadata aggregation, while the resulting audio and metadata features are fused by PCVG before classification.}
    \label{fig:framework}
\end{figure*}

\section{Proposed Method}
\label{sec:method}

\subsection{Overall Architecture}
\label{sec:architecture}

As illustrated in Fig.~\ref{fig:framework}, we adopt pretrained LAION-CLAP~\cite{clap,laionclap} as the backbone, using its HTSAT audio encoder and shared RoBERTa text encoder to obtain $\mathbf e_a$ and one $\mathbf e_v$ for each title, tag, or description field. A binary mask $\mathbf m=(m_v)_{v\in\mathcal V}$, where $\mathcal V$ is the field set and $m_v\in\{0,1\}$, excludes missing fields.

While the pretrained CLAP space aligns audio and text, it neither accounts for the varying relevance of heterogeneous metadata fields nor resolves dimension-wise cross-modal conflicts. To address this, FAM-Gate operates in two sequential stages: MVA first dynamically weighs and aggregates valid metadata representations, and PCVG subsequently regulates how the aggregated semantics interact with acoustic evidence along individual feature dimensions prior to classification.

\begin{figure*}[htb]
    \centering
    \includegraphics[width=0.8\textwidth]{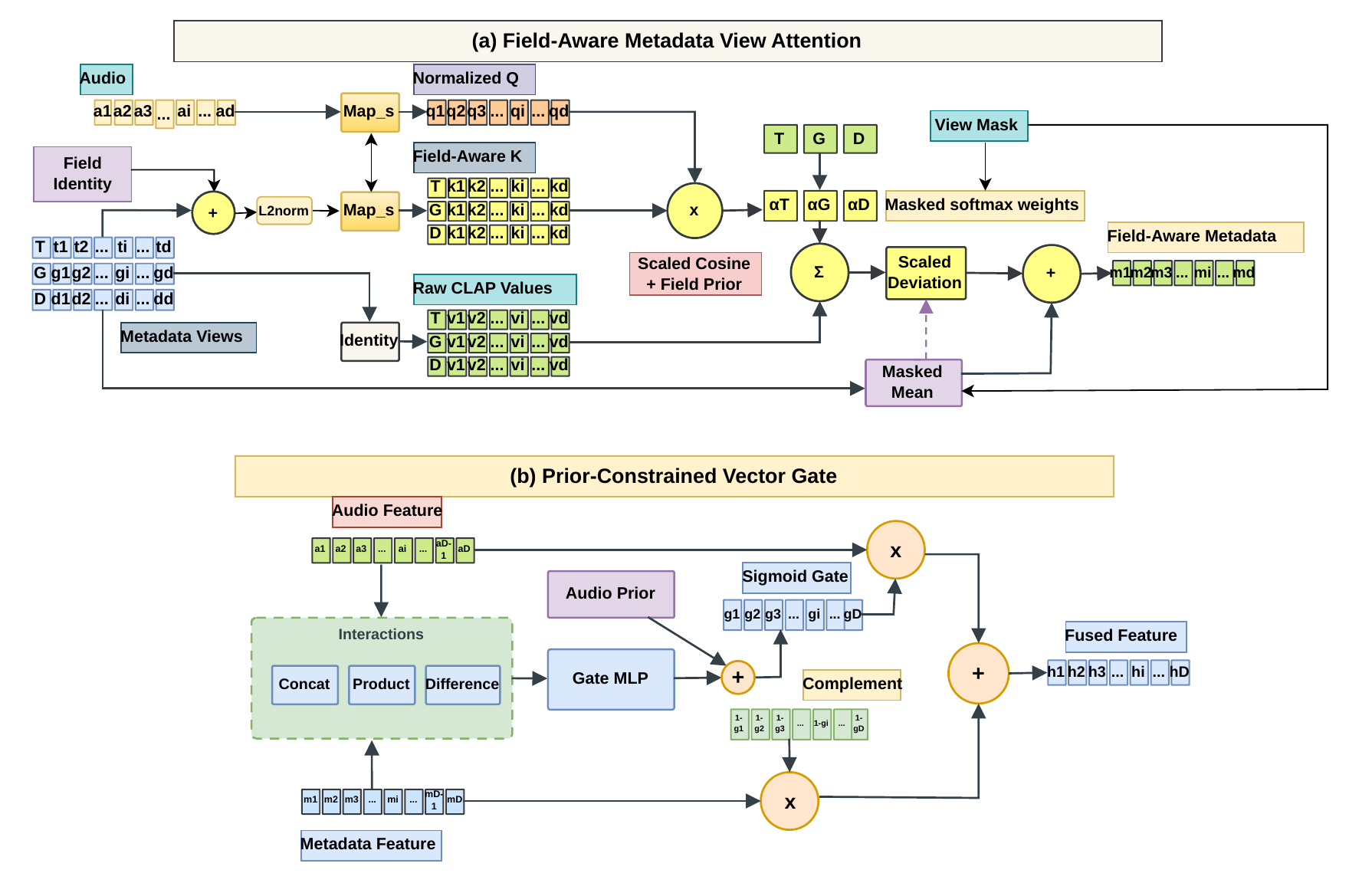}
    \caption{Detailed architectural diagrams of (a) MVA in Eqs.~\eqref{eq:mva_projection}--\eqref{eq:mva_output}, depicting the shared projection map $\mathcal S$, field prior $c_v$, view weights $\boldsymbol\alpha$, and mean-anchor deviation coefficient $\gamma$; and (b) PCVG in Eqs.~\eqref{eq:pcvg}--\eqref{eq:fusion}, highlighting the residual adjustment $\boldsymbol\delta$, acoustic prior $\boldsymbol\beta$, and gate vector $\mathbf g$. Post-fusion classification operations are omitted for clarity.}
    \label{fig:modules}
\end{figure*}

\subsection{Field-Aware Metadata View Attention}

Titles, tags, and descriptions present distinct trade-offs in structure, semantic density, and noise. Simple strategies like random selection drop valuable information, while uniform averaging assumes equal relevance across all fields. Since all metadata fields describe the same recording, the audio feature serves as a natural anchor for field selection. However, projecting queries and keys via independent linear transformations risks disrupting the geometry of the pretrained CLAP space. We therefore adopt a shared residual projection scheme, as shown in Fig.~\ref{fig:modules}(a):
\begin{equation}
\begin{aligned}
\mathcal{S}(\mathbf{z})&=\mathbf{z}+\mathbf{W}\mathbf{z},\\
\mathbf{q}&=\operatorname{norm}_2(\mathcal{S}(\mathbf{e}_a)),\\
\mathbf{k}_v&=\operatorname{norm}_2
\left(\mathcal{S}(\mathbf{e}_v)+\mathbf{p}_v\right),
\end{aligned}
\label{eq:mva_projection}
\end{equation}
where $\mathbf W$ is a shared learnable bias-free linear map and $\operatorname{norm}_2(\mathbf{x})=\mathbf{x}/\max(\|\mathbf{x}\|_2,\epsilon)$ denotes $\ell_2$ normalization with a small constant $\epsilon>0$ to prevent division by zero. This residual formulation allows zero/identity initialization, transforming feature coordinates in a shared space rather than learning separate audio and text mappings. The learnable embedding $\mathbf p_v$ preserves field identity under this unified transformation.

With normalized queries and keys, their inner product yields cosine audio--field relevance. We complement this sample-dependent relevance with a dataset-level field prior and the mask $\mathbf m$:
\begin{equation}
\begin{aligned}
z_v&=s\mathbf q^{\top}\mathbf k_v+c_v,\\
\alpha_v&=
\frac{m_v\exp(z_v)}
{\sum_{u\in\mathcal V}m_u\exp(z_u)},\qquad v\in\mathcal V,
\end{aligned}
\label{eq:mva_weights}
\end{equation}
where $z_v$, $s>0$, $c_v$, and $\exp$ denote the field score, learnable scale, field prior, and natural exponential, respectively. Assuming at least one field is available, $m_v$ filters out missing views while $\alpha_v$ normalizes the score across available fields rather than token sequences~\cite{transformer,multimodaltransformer}.

Softmax attention may become overly concentrated on a noisy metadata field. To mitigate this and preserve stable CLAP semantics, MVA retains the original text embeddings as values and anchors their aggregation to the masked mean representation:
\begin{equation}
\begin{aligned}
\mathbf{e}_{\mathrm{att}}&=\sum_{v\in\mathcal V}\alpha_v\mathbf{e}_v,\quad
\mathbf{e}_{\mathrm{mean}}=
\frac{\sum_{v\in\mathcal V}m_v\mathbf{e}_v}
{\max(1,\sum_{v\in\mathcal V}m_v)},\\
\gamma&=\tanh(\rho),\quad
\mathbf{e}_m=\mathbf{e}_{\mathrm{mean}}+
\gamma(\mathbf{e}_{\mathrm{att}}-\mathbf{e}_{\mathrm{mean}}).
\end{aligned}
\label{eq:mva_output}
\end{equation}
Here, $\rho$ is an unconstrained learnable scalar and $\tanh(\cdot)$ is the hyperbolic tangent activation, placing the trade-off factor $\gamma \in (-1,1)$ to bound the deviation from the field mean. Consequently, MVA grounds feature fusion around a stable mean anchor while dynamically applying audio-conditioned corrections; setting $\gamma=0$ gracefully recovers the exact masked mean baseline.

\subsection{Prior-Constrained Vector Gate}

Aggregating metadata fields resolves intra-text heterogeneity, but the interaction between aggregated semantic and acoustic evidence remains unresolved. Audio features emphasize spectro-temporal patterns and timbre, whereas metadata representations capture abstract categorical and contextual cues. Because these distinct modal properties can either complement or contradict each other across specific channels, a single scalar gate lacks the capacity to capture channel-wise variations. Given $\mathbf h_a,\mathbf h_m\in\mathbb R^D$, PCVG predicts a feature-wise gate based on both individual modal representations and their explicit interaction terms:
\begin{equation}
\begin{aligned}
\mathbf{r}&=[\mathbf{h}_a;\mathbf{h}_m;
\mathbf{h}_a\odot\mathbf{h}_m;
|\mathbf{h}_a-\mathbf{h}_m|],\\
\boldsymbol{\delta}&=\tanh
\left(\operatorname{MLP}(\operatorname{LN}(\mathbf{r}))\right),\\
\mathbf{g}&=\sigma(\boldsymbol{\beta}+\eta\boldsymbol{\delta}),
\end{aligned}
\label{eq:pcvg}
\end{equation}
following generic multimodal feature-wise fusion principles~\cite{multimodalsurvey,tensorfusion,film}. $\odot$ denotes the element-wise Hadamard product and $|\cdot|$ represents the element-wise absolute value. The combined vector $\mathbf r$ retains modality-specific evidence, while its Hadamard product and absolute difference terms explicitly model inter-modal agreement and conflict, respectively. $\operatorname{LN}$ denotes Layer Normalization, and $\operatorname{MLP}$ is a two-layer perceptron configured with Leaky-ReLU and dropout between its affine transformations. The outer $\tanh$ bounds the sample-dependent residual adjustment $\boldsymbol\delta$. Since audio is the primary observed modality while user-generated metadata can be noisy, a learnable vector $\boldsymbol\beta\in\mathbb R^D$ provides a baseline acoustic prior, with $\eta$ acting as a scaling factor to constrain the maximum dynamic range of the update. Finally, $\sigma(x)=(1+\exp(-x))^{-1}$ maps the gate logits into $\mathbf g\in(0,1)^D$.

The predicted vector gate performs dimension-wise feature interpolation:
\begin{equation}
\mathbf{h}_f=
\operatorname{Dropout}\!\left(
\operatorname{LN}\!\left(
\mathbf{g}\odot\mathbf{h}_a+
(\mathbf{1}_D-\mathbf{g})\odot\mathbf{h}_m
\right)\right),
\label{eq:fusion}
\end{equation}
where $\mathbf{1}_D\in\mathbb R^D$ is an all-ones vector and $\operatorname{Dropout}$ applies standard dropout regularization. As depicted in Fig.~\ref{fig:modules}(b), a larger gating value $g_i$ prioritizes acoustic evidence in dimension $i$, whereas a smaller value relies more heavily on semantic metadata. Layer Normalization stabilizes feature variance across channels prior to passing $\mathbf{h}_f$ into the residual classification head.

\subsection{Prediction and Objectives}

For a batch of $B$ recordings evaluated across $C$ target classes, let $o_{n,c}$ denote the predicted logit for sample $n$ and class $c$. For single-label classification datasets such as BSD, where each recording is annotated with a ground-truth label $y_n \in \{1,\ldots,C\}$, we optimize the standard categorical cross-entropy loss:
\begin{equation}
\mathcal{L}_{\mathrm{SL}}
=-\frac{1}{B}\sum_{n=1}^{B}
\ln
\frac{\exp(o_{n,y_n})}
{\sum_{c=1}^{C}\exp(o_{n,c})}.
\label{eq:single_label_loss}
\end{equation}

For multi-label settings like FSD50K, where a recording can be assigned multiple binary targets $y_{n,c} \in \{0,1\}$, we apply independent sigmoid activations and minimize the multi-label binary cross-entropy loss with logits:
\begin{equation}
\begin{aligned}
\mathcal{L}_{\mathrm{ML}}
=& -\frac{1}{BC}\sum_{n=1}^{B}\sum_{c=1}^{C}
\Big[
y_{n,c}\ln \sigma(o_{n,c})+ \\
& (1-y_{n,c})\ln\bigl(1-\sigma(o_{n,c})\bigr)
\Big].
\end{aligned}
\label{eq:multilabel_loss}
\end{equation}

\begin{table}[h]
\centering
\caption{Comparison with representative published systems on the official FSD50K evaluation set. Input modalities: A (Audio), M (Uploader-provided metadata).}
\label{tab:fsd_comparison}
\setlength{\tabcolsep}{4pt}
\renewcommand{\arraystretch}{1.1}
\scriptsize
\resizebox{\columnwidth}{!}{%
\begin{tabular}{lcccc}
\hline
Method & Venue & Pretraining & Input & mAP (\%) \\
\hline
PaSST-S~\cite{passt}
& Interspeech'22 & Img.+AudioSet & A & 65.30 \\
TFECN~\cite{tfecn}
& Interspeech'23 & Img.+AudioSet & A & 67.29 \\
MN ($\alpha=3.0$)~\cite{dynamiccnn}
& TASLP'24 & AudioSet & A & 65.60 \\
DyMN-L~\cite{dynamiccnn}
& TASLP'24 & AudioSet & A & 65.50 \\
\hline
\textbf{FAM-Gate}
& Ours & LAION-CLAP & A+M & \textbf{74.98$\pm$0.23} \\
\hline
\end{tabular}%
}
\end{table}

\section{Experiments}
\label{sec:experiments}

\subsection{Datasets and Experimental Settings}

\begin{table*}[!t]
    \centering
    \caption{Ablation study across five independent random seeds under a unified evaluation protocol (\%, mean $\pm$ standard deviation). FAM-Gate incorporates the proposed Field-Aware MVA with PCVG.}
    \label{tab:main_results}
    \footnotesize
    \setlength{\tabcolsep}{3.0pt}
    \renewcommand{\arraystretch}{1}
    \begin{tabular}{lcccccc}
        \toprule
        & \multicolumn{2}{c}{BSD10K}
        & \multicolumn{2}{c}{BSD35K}
        & \multicolumn{2}{c}{FSD50K} \\
        \cmidrule(lr){2-3}
        \cmidrule(lr){4-5}
        \cmidrule(lr){6-7}
        Method & Acc. & hF & Acc. & hF & mAP & Macro-F1 \\
        \midrule

        Random View + Scalar
        & $79.49\pm0.30$ & $77.76\pm0.76$
        & $83.43\pm0.16$ & $75.90\pm2.46$
        & $69.90\pm0.99$ & $61.44\pm2.52$ \\

        PCVG Only
        & $80.06\pm0.24$ & $78.60\pm0.60$
        & $83.62\pm0.26$ & $75.77\pm1.08$
        & $71.99\pm0.82$ & $64.68\pm2.82$ \\

        Mean + PCVG
        & $81.04\pm0.64$ & $79.57\pm1.55$
        & $\mathbf{87.59\pm0.24}$ & $79.79\pm2.14$
        & $74.88\pm0.39$ & $71.18\pm1.02$ \\

        Projected MVA + Scalar
        & $80.22\pm0.63$ & $79.02\pm0.56$
        & $86.62\pm0.15$ & $79.12\pm1.75$
        & $71.77\pm1.43$ & $67.09\pm2.26$ \\

        Projected MVA + PCVG
        & $81.19\pm0.40$ & $79.76\pm0.80$
        & $87.29\pm0.34$ & $80.30\pm1.61$
        & $74.86\pm0.26$ & $71.17\pm0.78$ \\

        \textbf{FAM-Gate}
        & $\mathbf{81.50\pm0.67}$ & $\mathbf{80.29\pm1.48}$
        & $87.51\pm0.22$ & $\mathbf{81.62\pm1.51}$
        & $\mathbf{74.98\pm0.23}$ & $\mathbf{71.69\pm0.87}$ \\

        \bottomrule
    \end{tabular}
\end{table*}

We evaluate our framework on three standard audio benchmarks: BSD10K, BSD35K, and FSD50K. BSD10K and BSD35K consist of 23 and 28 single-label leaf categories, respectively. For both BSD variants, we use a fixed 90/10 sample-level split generated with seed 3407~\cite{bsd2024,bsd2025}, reporting leaf accuracy and weighted hierarchical F-score (hF) with $\lambda=0.75$. FSD50K contains 200 multi-label sound classes, and we follow its official train, validation, and evaluation partitions~\cite{fsd50k}. Performance on FSD50K is measured using macro mAP and macro F1 at a decision threshold of 0.5.

To isolate component contributions, we evaluate six matched variants. Random View + Scalar, adapted from~\cite{bsd2025}, randomly selects one available field and applies scalar audio--text fusion. PCVG Only isolates dimension-wise fusion by replacing the scalar gate, while Mean + PCVG tests the benefit of retaining all valid fields. Projected MVA + Scalar and Projected MVA + PCVG use the earlier attention design with independent query/key/value projections and projected text values, paired with scalar fusion and PCVG, respectively. FAM-Gate combines Field-Aware MVA with PCVG.

All six variants share the same LAION-CLAP backbone, prediction head, dataset splits, and training hyper-parameters. Each configuration is evaluated with five fixed random seeds, and we report the mean $\pm$ sample standard deviation. Each run is trained on a single NVIDIA RTX A6000 GPU.

\subsection{Implementation Details}

We initialize the HTSAT-tiny audio backbone and RoBERTa text encoder using the official 630k-AudioSet LAION-CLAP checkpoint, fine-tuning the full architecture end-to-end. Audio inputs are processed via standard CLAP preprocessing, incorporating a $48\text{ kHz}$ sampling rate, 10s fusion truncation, and repeat padding. Titles, uploader tags, and descriptions are encoded independently, with unavailable fields explicitly masked out. To prevent label leakage, ground-truth labels and class descriptions are withheld from the text encoder during training.

We optimize with Adam using a batch size of 64, weight decay of $10^{-5}$, and dropout of 0.1. Learning rates are $5\times10^{-6}$ for CLAP and $10^{-4}$ for task-specific modules. Training runs for at most 100 epochs; StepLR halves both rates every 10 epochs, and early stopping uses a patience of 15. The MVA temperature is initialized to 1, while PCVG uses $\eta=0.5$ and an initial audio prior of 0.65.

\subsection{Overall Performance}

Table~\ref{tab:fsd_comparison} provides cross-setting context against representative published systems on the official 200-class FSD50K evaluation set. The listed systems use audio alone at inference, whereas FAM-Gate additionally uses uploader-provided metadata. FAM-Gate obtains $74.98\%$ mAP, numerically exceeding the highest listed audio-only result by $7.69$ percentage points. Because the input modalities differ, this is not a strictly modality-matched ranking; controlled evidence for the proposed modules is provided in Table~\ref{tab:main_results}.

\subsection{Ablation Study}

Table~\ref{tab:main_results} presents matched ablations across five independent runs. FAM-Gate achieves the highest five-seed mean on five of six metrics and consistently outperforms Random View + Scalar. It improves BSD10K accuracy/hF by 2.01/2.53 percentage points, BSD35K accuracy/hF by 4.09/5.72 points, and FSD50K mAP/macro F1 by 5.07/10.25 points. The notable gains in hF and macro F1 indicate that performance improvements are particularly pronounced under hierarchy-sensitive and class-balanced evaluation.

Stepwise comparisons isolate the individual contributions of each module. PCVG Only improves FSD50K mAP/macro F1 by 2.09/3.24 points over the scalar baseline, supporting the benefit of dimension-wise fusion. Retaining all available metadata fields (Mean + PCVG) yields further gains across tasks. With PCVG fixed, FAM-Gate raises BSD10K/BSD35K hF over Mean + PCVG by 0.72/1.83 points and FSD50K macro F1 by 0.51 points. While its BSD35K accuracy is 0.08 points lower than Mean + PCVG, its hF is 1.83 points higher, reflecting improved hierarchical consistency without sacrificing leaf-level precision. FAM-Gate also achieves higher five-seed means than Projected MVA + PCVG across all six metrics, supporting the design of our alignment-preserving, field-aware MVA.

Regarding complexity, the baseline and FAM-Gate contain 165.668M
and 165.982M parameters, respectively, an increase of only 0.314M
(0.19\%). Excluding CLAP encoders, task-side computation rises from
7.286M to 8.386M MACs per sample. These counts exclude the additional
cost of encoding three metadata views instead of one.

\begin{figure}[htb]
    \centering
    \includegraphics[width=\columnwidth]{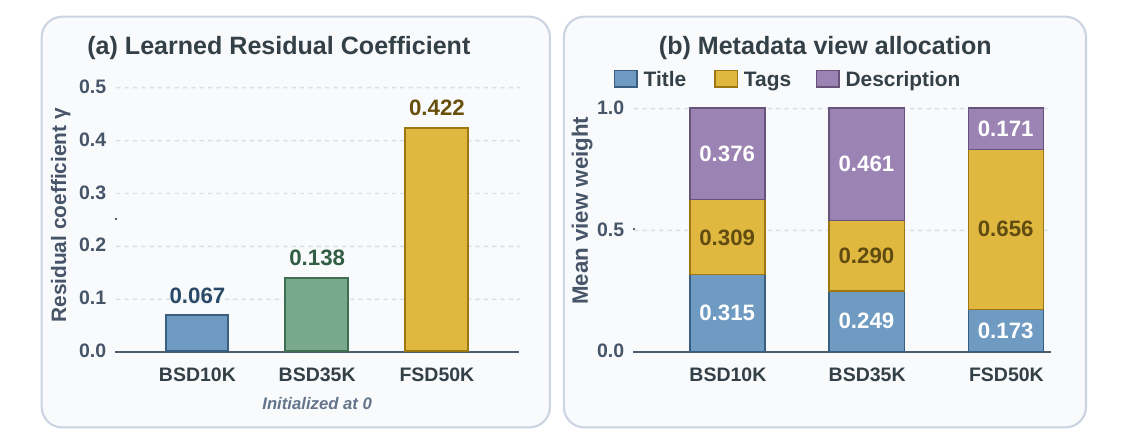}
    \caption{Learned metadata behavior aggregated over the five best checkpoints per dataset: (a) mean-anchor deviation coefficient $\gamma$, where $\gamma=0$ recovers masked mean aggregation, and (b) average sample-wise field weights.}
    \label{fig:metadata_behavior}
\end{figure}

Figure~\ref{fig:metadata_behavior} examines whether MVA performs
adaptive field aggregation rather than remaining close to a fixed
mean. Since $\gamma=0$ recovers masked mean aggregation, the learned
values of 0.067, 0.138, and 0.422 indicate increasingly stronger
attention-based adjustments on BSD10K, BSD35K, and FSD50K,
respectively. On BSD35K, description receives the highest average
weight (0.461), and masking it reduces hF by 6.84 points. On the
FSD50K validation set, tags receive an average weight of 0.656;
removing them on the evaluation subset reduces macro F1 by 5.12
points. Across-sample standard deviations of 0.040--0.101 further
show that the field weights vary across recordings. Because weight
rankings and masking effects do not fully agree on BSD10K, we
interpret attention weights as adaptive allocation patterns rather
than causal importance scores.

\section{Conclusion}
We introduced FAM-Gate, employing field-aware metadata aggregation
with prior-constrained vector fusion. Five-seed experiments across
three datasets demonstrate consistent improvements in hierarchical
and macro-level metrics over matched baselines. Future work will
address missing or corrupted metadata and broader audio-language
backbones.

\clearpage

\bibliographystyle{IEEEbib}
\bibliography{IEEEabrv, myabrv, refs}

\end{document}